\documentclass[12pt]{article}
\usepackage{graphics}
\usepackage{enumerate}
\usepackage{amssymb,epsfig,amsmath,euscript,array}
\usepackage{axodraw}
\usepackage{subfigure}
\usepackage{color}
\usepackage[sort&compress,square]{natbib}

\makeatletter
\@addtoreset{equation}{section}
\makeatother

\definecolor{blue}{rgb}{0,0,1}

\definecolor{red}{rgb}{1,0,0}

\definecolor{green}{rgb}{0,1,0}

\newcounter{multieqs}

\newcommand{\be}{\begin{equation}}
\newcommand{\ee}{\end{equation}}

\newcommand{\bm}[1]{\mbox{\boldmath $#1$}}

\def\bd{\begin{document}}
\def\ed{\end{document}}
\def\nn{\nonumber}
\def\bea{\begin{eqnarray}}
\def\eea{\end{eqnarray}}
\let\bm=\bibitem
\let\la=\label

\newcommand{\EQ}[1]{\begin{equation} #1 \end{equation}}
\newcommand{\AL}[1]{\begin{subequations}\begin{align} #1 \end{align}\end{subequations}}
\newcommand{\SP}[1]{\begin{equation}\begin{split} #1 \end{split}\end{equation}}
\newcommand{\ALAT}[2]{\begin{subequations}\begin{alignat}{#1} #2 \end{alignat}\end{subequations}}
\def\beqa{\begin{eqnarray}}
\def\eeqa{\end{eqnarray}}
\def\beq{\begin{equation}}
\def\eeq{\end{equation}}

\def\hf{{\textstyle \frac{1}{2}}}
\def\wbar{\bar w}
\def\mubar{\bar\mu}
\def\abar{\bar a}
\def\sigmabar{\bar\sigma}
\def\etabar{\bar\eta}
\def\zetabar{\bar\zeta}
\def\mubar{\bar\mu}
\def\nubar{\bar\nu}
\def\N{{\cal N}}
\def\sst{\scriptscriptstyle}
\def\thetabar{\bar\theta}
\def\Tr{{\rm Tr}}
\def\one{\mbox{1 \kern-.59em {\rm l}}}
 \def\Nh{\hat{N}}

\newlength{\myVSpace}
\newcommand\xstrut{\raisebox{-.5\myVSpace}
  {\rule{0pt}{\myVSpace}}%
}

\def\a{\alpha}      \def\da{{\dot\alpha}}
\def\b{\beta}       \def\db{{\dot\beta}}
\def\c{\gamma}  \def\G{\Gamma}  \def\cdt{\dot\gamma}
\def\d{\delta}  \def\D{\Delta}  \def\ddt{\dot\delta}
\def\e{\epsilon}        \def\vare{\varepsilon}
\def\f{\phi}    \def\F{\Phi}    \def\vvf{\f}
\def\h{\eta}
\def\k{\kappa}
\def\l{\lambda} \def\L{\Lambda}
\def\m{\mu} \def\n{\nu}
\def\o{\omega}
\def\p{\pi} \def\P{\Pi}
\def\r{\rho}
\def\s{\sigma}  \def\S{\Sigma}
\def\t{\tau}
\def\th{\theta} \def\Th{\Theta} \def\vth{\vartheta}
\def\X{\Xeta}
\def\z{\zeta}

\def\cA{{\cal A}} \def\cB{{\cal B}} \def\cC{{\cal C}}
\def\cD{{\cal D}} \def\cE{{\cal E}} \def\cF{{\cal F}}
\def\cG{{\cal G}} \def\cH{{\cal H}} \def\cI{{\cal I}}
\def\cJ{{\cal J}} \def\cK{{\cal K}} \def\cL{{\cal L}}
\def\cM{{\cal M}} \def\cN{{\cal N}} \def\cO{{\cal O}}
\def\cP{{\cal P}} \def\cQ{{\cal Q}} \def\cR{{\cal R}}
\def\cS{{\cal S}} \def\cT{{\cal T}} \def\cU{{\cal U}}
\def\cV{{\cal V}} \def\cW{{\cal W}} \def\cX{{\cal X}}
\def\cY{{\cal Y}} \def\cZ{{\cal Z}}

\def\ua{\underline{\alpha}}
\def\ub{\underline{\phantom{\alpha}}\!\!\!\beta}
\def\uc{\underline{\phantom{\alpha}}\!\!\!\gamma}
\def\um{\underline{\mu}}
\def\ud{\underline\delta}
\def\ue{\underline\epsilon}
\def\una{\underline a}\def\unA{\underline A}
\def\unb{\underline b}\def\unB{\underline B}
\def\unc{\underline c}\def\unC{\underline C}
\def\und{\underline d}\def\unD{\underline D}
\def\une{\underline e}\def\unE{\underline E}
\def\unf{\underline{\phantom{e}}\!\!\!\! f}\def\unF{\underline F}
\def\unm{\underline m}\def\unM{\underline M}
\def\unn{\underline n}\def\unN{\underline N}
\def\unp{\underline{\phantom{a}}\!\!\! p}\def\unP{\underline P}
\def\unq{\underline{\phantom{a}}\!\!\! q}
\def\unQ{\underline{\phantom{A}}\!\!\!\! Q}
\def\unH{\underline{H}}

\def\As {{A \hspace{-6.4pt} \slash}\;}
\def\bs {{b \hspace{-6.4pt} \slash}\;}
\def\Ds {{D \hspace{-6.4pt} \slash}\;}
\def\ds {{\del \hspace{-6.4pt} \slash}\;}
\def\ss {{\s \hspace{-6.4pt} \slash}\;}
\def\ks {{ k \hspace{-6.4pt} \slash}\;}
\def\ps {{p \hspace{-6.4pt} \slash}\;}
\def\pas {{{p_1} \hspace{-6.4pt} \slash}\;}
\def\pbs {{{p_2} \hspace{-6.4pt} \slash}\;}

\def\Fh{\hat{F}}
\def\Vh{\hat{V}}
\def\Xh{\hat{X}}
\def\ah{\hat{a}}
\def\xh{\hat{x}}
\def\yh{\hat{y}}
\def\ph{\hat{p}}
\def\xih{\hat{\xi}}

\def\psit{\tilde{\psi}}
\def\Psit{\tilde{\Psi}}
\def\tht{\tilde{\th}}

\def\At{\tilde{A}}
\def\Qt{\tilde{Q}}
\def\Rt{\tilde{R}}
\def\Nt{\tilde{N}}

\def\at{\tilde{a}}
\def\st{\tilde{s}}
\def\ft{\tilde{f}}
\def\pt{\tilde{p}}
\def\qt{\tilde{q}}
\def\vt{\tilde{v}}
\def\nt{\tilde{n}}

\def\delb{\bar{\partial}}
\def\bz{\bar{z}}
\def\bD{\bar{D}}
\def\bB{\bar{B}}

\def\bk{{\bf k}}
\def\bl{{\bf l}}
\def\bp{{\bf p}}
\def\bq{{\bf q}}
\def\br{{\bf r}}
\def\bx{{\bf x}}
\def\by{{\bf y}}
\def\bR{{\bf R}}
\def\bV{{\bf V}}

\def\d{\delta}\def\D{\Delta}\def\ddt{\dot\delta}

\def\pa{\partial} \def\del{\partial}
\def\xx{\times}
\def\uno{\mbox{1 \kern-.59em {\rm l}}}

\def\trp{^{\top}}
\def\inv{^{-1}}
\def\dag{{^{\dagger}}}
\def\pr{^{\prime}}

\def\rar{\rightarrow}
\def\lar{\leftarrow}
\def\lrar{\leftrightarrow}

\newcommand{\0}{\,\!}      
\def\one{1\!\!1\,\,}
\def\im{\imath}
\def\jm{\jmath}

\newcommand{\tr}{\mbox{tr}}
\newcommand{\slsh}[1]{/ \!\!\!\! #1}

\def\vac{|0\rangle}
\def\lvac{\langle 0|}

\def\hlf{\frac{1}{2}}
\def\ove#1{\frac{1}{#1}}

\def\Box{\square}
\def\ZZ{\mathbb{Z}}
\def\CC#1{({\bf #1})}
\def\bcomment#1{}
\def\bfhat#1{{\bf \hat{#1}}}
\def\VEV#1{\left\langle #1\right\rangle}
\def\vev#1{\langle{#1}\rangle}

\newcommand{\ex}[1]{{\rm e}^{#1}} \def\ii{{\rm i}}

\def\rr{{\rm r}} \def\rs{{\rm s}}\def\rv{{\rm v}}
\def\ri{{\rm i}}\def\rj{{\rm j}}

\newcommand{\lrbrk}[1]{\left(#1\right)}
\newcommand{\sfrac}[2]{{\textstyle\frac{#1}{#2}}}

\font\mybb=msbm10 at 12pt
\def\bb#1{\hbox{\mybb#1}}

\font\myBB=msbm10 at 18pt
\def\BB#1{\hbox{\myBB#1}}

\begin{document}
\noindent
\hspace*{13.5cm}DESY 05-127\\
\hspace*{13.5cm}IPPP/05/45\\
\hspace*{13.5cm}DCPT/05/90

\vspace{25pt}

\begin{center}

{\Large \bf  Telltale Traces of U(1) Fields\\
in Noncommutative Standard Model Extensions\\[1.5ex]
}

\vspace{30pt}

{\bf Joerg Jaeckel$^1$, Valentin V.  Khoze$^2$ and Andreas Ringwald$^1$}

{\small \em
{}$^1$Deutsches Elektronen-Synchrotron DESY,
Notkestrasse 85, D-22607  Hamburg, Germany\\
{}$^2$Department of Physics and IPPP, University of Durham,
Durham, DH1 3LE, UK

\vspace{10pt}

{\sffamily \tt
joerg.jaeckel@desy.de, valya.khoze@durham.ac.uk, andreas.ringwald@desy.de}
}

\vspace{30pt}
{\bf Abstract}
\end{center}

\noindent {Restrictions imposed by gauge invariance in noncommutative spaces
together with the effects of ultraviolet/infrared mixing lead to strong constraints
on possible candidates for a noncommutative extension of the Standard Model.
In this paper, we study a general class of
4-dimensional 
noncommutative models consistent with these restrictions.
Specifically we consider models 
based upon a gauge theory with the gauge group
${\textrm U}(N_1)\times {\textrm U}(N_2) \times \ldots \times {\textrm U}(N_m)$
coupled to matter fields transforming in
the (anti)-fundamental, bi-fundamental and adjoint representations.
Noncommutativity is introduced using the Weyl-Moyal star-product approach on a continuous space-time.
We pay particular attention to overall trace-U(1) factors of the gauge group which
are affected by the ultraviolet/infrared mixing. We show that, in general, these trace-U(1)
gauge fields do not decouple sufficiently fast in the infrared, and lead to sizable
Lorentz symmetry violating effects in the low-energy effective theory. Making these
effects unobservable
in the class of models we consider 
would require pushing the
constraint on the noncommutativity mass scale far beyond the Planck
mass ($M_{\textrm{NC}}\gtrsim 10^{100}\, M_{\textrm{P}}$) and
severely limits the phenomenological prospects of such models.}
\setcounter{page}{0} \thispagestyle{empty}

\newpage

\section{Introduction and discussion of results}

Gauge theories on spaces with noncommuting coordinates
\begin{equation}
[x^\mu,x^\nu]=i\,\theta^{\mu\nu} \ ,
\end{equation}
provide a very interesting
new class of quantum field theories with intriguing and sometimes unexpected features.
These noncommutative models can arise naturally as low-energy effective theories from string
theory and D-branes. As field theories they must satisfy a number of restrictive
constraints detailed below, and this makes them particularly interesting and challenging
for purposes of particle physics model building.
For general reviews of noncommutative gauge theories the reader can consult e.g.
Refs.~\cite{Seiberg:1999vs,Douglas:2001ba,Szabo:2001kg}.

There are two distinct approaches used in the recent literature for constructing
quantum field theories on noncommutative spaces. The first approach uses the Weyl-Moyal
star-products to introduce noncommutativity. In this case,
noncommutative field theories are defined by replacing the ordinary
products of all fields in the Lagrangians of their commutative counterparts
by the star-products
\begin{equation}
(\phi * \varphi) (x) \equiv \phi(x)\  e^{{i\over 2}\theta^{\mu\nu}
\stackrel{\leftarrow}{\partial_\mu}
\stackrel{\rightarrow}{\partial_\nu}} \  \varphi(x) \ . \label{stardef}
\end{equation}
Noncommutative theories in the Weyl-Moyal formalism can be viewed as field theories on
 ordinary commutative spacetime. For example,  the noncommutative pure
gauge theory action is
\begin{equation}
S = -{1\over 2g^2}\int d^{4} x \ \Tr ( F_{\mu \nu}*  F^{\mu \nu}
 ) \ , \label{pureym}
\end{equation}
where
the commutator in the field strength also contains the star-product. The important feature
of this approach is the fact that phase factors in the star-products are not expanded in powers of $\theta$ and
the $\theta$ dependence in the Lagrangian is captured entirely. This ability to work to all orders in $\theta$
famously gives rise to the ultraviolet/infrared (UV/IR) mixing ~\cite{Minwalla:1999px,Matusis:2000jf}
in the noncommutative quantum field theory which we will review below.

The second approach to noncommutativity does not employ star-products. It instead
relies \cite{Madore:2000en,Calmet:2001na}
 on the Seiberg-Witten map which represents noncommutative fields as
a function of $\theta$ and ordinary commutative fields.
This approach essentially reduces noncommutativity to an introduction
of an infinite set of higher-dimensional (irrelevant) operators, each suppressed
by the corresponding power of $\theta$, into the action.
There are two main differences compared to the Weyl-Moyal approach. First, in practice one always works with the first few terms in the power series in $\theta$ and in this setting the UV/IR mixing cannot be captured. Second,
the Seiberg-Witten map is a non-linear field transformation. Therefore, one expects a non-trivial Jacobian and possibly a quantum theory different from the one obtained in the Weyl-Moyal approach.
In the
rest of this paper we will concentrate on the Weyl-Moyal approach.

In the context of Weyl-Moyal
noncommutative Standard
Model building, a number of features of noncommutative
gauge theories have to be taken into account which are believed
to be generic~\cite{Khoze:2004zc}:
\begin{enumerate}[1.\,]
\item{} the mixing of ultraviolet and infrared effects~\cite{Minwalla:1999px,Matusis:2000jf}
        and the asymptotic decoupling of U(1) degrees of freedom~\cite{Khoze:2000sy,Hollowood:2001ng} in the
        infrared;
\item{} the gauge groups are restricted to U($N$) groups~\cite{Matsubara:2000gr,Armoni:2000xr} or products of thereof;
\item{} fields can transform only in (anti-)fundamental, bi-fundamental
        and adjoint representations~\cite{Gracia-Bondia:2000pz,Terashima:2000xq,Chaichian:2001mu};
\item{} the charges of matter fields are restricted~\cite{Hayakawa:1999zf}
        to $0$ and $\pm 1$,
        thus requiring extra care in order to give fractional electric charges
        to the quarks;
\item{} gauge anomalies cannot be cancelled in a chiral noncommutative
        theory~\cite{Hayakawa:1999zf,Ardalan:2000cy,Gracia-Bondia:2000pz,Bonora:2000he,Martin:2000qf,Intriligator:2001yu,Armoni:2002fh},
        hence the anomaly-free gauge theory must be
        vector-like.
\end{enumerate}

Building upon an earlier proposal by Chaichian {\it{et al.}} \cite{Chaichian:2001py},
the authors of Ref.~\cite{Khoze:2004zc}
constructed an example of a noncommutative
embedding of the Standard Model with the purpose to satisfy all the requirements listed above.
The model of \cite{Khoze:2004zc} is based on the gauge group $\textrm{U}(4)\times \textrm{U}(3) \times \textrm{U}(2)$
with matter fields transforming
in noncommutatively allowed representations. Higgs fields break the noncommutative gauge group down to
a low-energy commutative gauge theory which includes
the Standard Model group $\textrm{SU}(3)\times \textrm{SU}(2) \times \textrm{U}(1)_Y$.
The $\textrm{U}(1)_Y$ group here corresponds to
ordinary QED, or more precisely to the hypercharge $Y$
Abelian gauge theory. The generator of $\textrm{U}(1)_Y$ was constructed from a linear combination of
{\it traceless} diagonal generators of the microscopic theory $\textrm{U}(4)\times \textrm{U}(3) \times \textrm{U}(2).$
Because of this, the UV/IR effects -- which can affect only the overall trace-$\textrm{U}(1)$ subgroup of
each $\textrm{U}(N)$ -- were not contributing to the hypercharge $\textrm{U}(1)_Y.$ However some of the
overall trace-$\textrm{U}(1)$ degrees of freedom
can survive the Higgs mechanism and thus contribute to the low-energy
effective theory, in addition to the Standard Model fields. These additional
trace-$\textrm{U}(1)$ gauge fields logarithmically decouple from the low-energy effective theory
and were neglected in the analysis of Ref.~\cite{Khoze:2004zc}.
The main goal of the present paper is to take these effects into account.

We will find that
the noncommutative model building constraints,
and, specifically, the UV/IR mixing effects in the trace-U(1) factors
in the item 1 above,
lead to an unacceptable defective behavior of the low-energy theory,
when we try to construct a model having the photon as the only
massless colourless U(1) gauge boson.
Our findings rule out a class of noncommutative extensions of the Standard Model.

(a) This class is based on a noncommutative quantum gauge theory defined on a four-dimensional continuous
space-time (UV cutoff sent to infinity). Within the Weyl-Moyal approach there are two ways to avoid our conclusions. Either one can introduce extra dimensions \cite{AJKR} or one can give up the continuous space-time. 

(b) Noncommutative models we concentrate on are similar to the example in~\cite{Khoze:2004zc}
and should be distinguished from earlier ones studied in \cite{Chaichian:2001py} for two
reasons.\footnote{
The construction in \cite{Khoze:2004zc} of correct values of
hypercharges of the Standard Model
from the product gauge group was influenced by \cite{Chaichian:2001py}.
The authors of Ref.~\cite{Chaichian:2001py} advocated a
noncommutative model which satisfied criteria 2, 3 and 4 listed in
the beginning of this section.
Their model was based on the noncommutative gauge group
$\textrm{U}(3)\times \textrm{U}(2) \times \textrm{U}(1)$ with matter fields transforming
only in (bi-)fundamental representations, and remarkably, it predicted correctly
the hypercharges of the Standard Model. In many respects their model
is similar to the bottom-up approach of \cite{bottomup}
to the string embedding of the Standard Model in purely commutative settings.
Unfortunately, the noncommutative $\textrm{U}(3)\times \textrm{U}(2) \times \textrm{U}(1)$ model
of \cite{Chaichian:2001py} ignores all the effects of the UV/IR mixing which
alters the infrared behavior of
the U(1) hypercharge sector.}
First, we include the effects of the UV/IR mixing in our analysis.
Second, is that our models preserve full noncommutative gauge invariance including the Higgs and Yukawa sectors.
As such, the difficulties related to unitarity violation discussed in \cite{Hewett:2001im} do not apply in our case.

(c) Finally, as already mentioned earlier, we are not pursuing the Seiberg-Witten map approach
and as such our conclusions cannot be directly applied to the class of noncommutative models
which rely on Taylor expansion in powers of $\theta$
in \cite{Calmet:2001na,Madore:2000en,Carroll:2001ws,
Carlson:2001sw,Behr:2002wx,Schupp:2002up,Calmet:2004dn,Ohl:2004tn,Melic:2005su}.

The UV/IR mixing in noncommutative theories arises from the fact that
certain classes of Feynman diagrams acquire factors of the form
$e^{i k_\mu \theta^{\mu\nu} p_\nu}$
(where $k$ is an external momentum and $p$ is a loop momentum) compared to their commutative
counter-parts.
These factors directly follow from the use of the Weyl-Moyal star-product \eqref{stardef}.
At large values of the loop momentum $p$,
the oscillations of $e^{i k_\mu \theta^{\mu\nu} p_\nu}$
improve
the convergence of the loop integrals. However, as the external momentum vanishes, $k \to 0,$
the divergence reappears and
what would have been a UV divergence is now reinterpreted as an IR divergence instead.
This phenomenon of UV/IR mixing is specific to noncommutative theories and
does not occur in the commutative settings where the physics of high energy degrees
of freedom does not affect the physics at low energies.

There are two important points concerning the UV/IR
mixing~\cite{Matusis:2000jf,Khoze:2000sy,Hollowood:2001ng,Armoni:2000xr}
which we want to stress here.
First, the UV/IR mixing occurs only in the trace-U(1) components of the
noncommutative $\textrm{U}(N)$ theory, leaving the $\textrm{SU}(N)$ degrees of freedom unaffected.
Second, there are two separate sources of the UV/IR mixing contributing to the
dispersion relation of the trace-U(1) gauge fields: the $\Pi_1$ effects
and the $\Pi_2$ effects, as will be explained momentarily.

A study of the Wilsonian effective action, obtained by integrating out the
high-energy degrees of freedom using the background field method,
and keeping track of the UV/IR mixing effects,
has given strong hints in favour of a non-universality in the infrared~\cite{Khoze:2000sy,Hollowood:2001ng}.
In particular, the polarisation tensor of
the gauge bosons in a noncommutative $\textrm{U}(N)$ gauge theory takes form
~\cite{Matusis:2000jf,Khoze:2000sy,Hollowood:2001ng}
\begin{equation}
\label{poltensor}
\Pi_{\mu\nu}^{AB} = \Pi_1^{AB}(k^2,\tilde k^2) \, \left( k^2 g_{\mu\nu} - k_\mu k_\nu \right)
+ \Pi_{2}^{AB} (k^2, \tilde k^2)\, \frac{\tilde{k}_{\mu}\tilde{k}_{\nu}}{\tilde{k}^2}
\,, \hspace{4ex} {\rm with\ } \tilde{k}_\mu = \theta_{\mu\nu} k^\nu \,.
\end{equation}
Here $A,B=0,1,\ldots N^2-1$ are adjoint labels of $\textrm{U}(N)$ gauge fields, $A_\mu^A$,
such that $A,B=0$ correspond to the overall $\textrm{U}(1)$ subgroup, i.e. to the
trace-U(1) factor.
The term in \eqref{poltensor}
proportional to $\tilde{k}_\mu\tilde{k}_\nu /\tilde{k}^2 $
would not appear in
ordinary commutative theories. It is transverse, but
not Lorentz invariant, as it explicitly depends on $\theta_{\mu\nu}.$
Nevertheless it is perfectly allowed in noncommutative theories.
It is known
that $\Pi_2$ vanishes
for supersymmetric noncommutative gauge theories with unbroken supersymmetry,
as was first discussed
in \cite{Matusis:2000jf}.

In general, both $\Pi_1$ and $\Pi_2$ terms in \eqref{poltensor} are affected by the
UV/IR mixing. More precisely, as already mentioned earlier,
the UV/IR mixing affects specifically the $\Pi_1^{0\,0}$ components and
generates the $\Pi_2^{0\,0}$ components
in \eqref{poltensor}.
The UV/IR mixing in $\Pi_1^{0\,0}$ affects the running of the trace-U(1) coupling
constant in the infrared,
\begin{equation}
\frac{1}{g(k,\tilde{k})_{\textrm{U}(1)}^2} = 4 \Pi_{1}^{0\,0}( k^2,\tilde k^2)
\, \rightarrow \,
-\,\frac{b_0}{(4\pi)^2} \, \log { k^2} \ , \qquad {\rm as} \
k^2\to 0
\,,
\end{equation}
leading to a logarithmic decoupling of the trace-U(1) gauge fields
from the $\textrm{SU}(N)$ low-energy theory, see
Refs.~\cite{Khoze:2004zc,Khoze:2000sy,Hollowood:2001ng} for more detail.

For nonsupersymmetric theories,
$\Pi_2^{0\,0}$ can present more serious problems.
In theories without supersymmetry, $\Pi_2^{0\,0} \sim 1/{\tilde{k}^2},$ at small momenta,
and this leads to unacceptable quadratic IR singularities \cite{Matusis:2000jf}.
In theories with softly broken supersymmetry (i.e. with matching number of
bosonic and fermionic degrees of freedom) the quadratic singularities in $\Pi_2^{0\,0}$
cancel~\cite{Matusis:2000jf,Khoze:2000sy,Hollowood:2001ng}. However, the subleading
contribution $\Pi_2^{0\,0} \sim const,$ survives \cite{Alvarez-Gaume:2003}
unless the supersymmetry is exact.
For the rest of the paper we will
concentrate on noncommutative Standard Model candidates with softly broken supersymmetry,
in order to avoid quadratic IR divergencies. In this case,
$\Pi_2^{0\,0} \sim \Delta M^2_{\rm susy},$\footnote{$\Delta M^2_{\rm SUSY}
=\frac{1}{2}\sum_s M_s^2-\sum_f M_f^2$
is a measure of SUSY breaking.} as explained in \cite{Alvarez-Gaume:2003}.
The presence of such $\Pi_2$ effects will lead to unacceptable pathologies such as
Lorentz-noninvariant dispersion
relations giving mass to only one of the polarisations of the trace-U(1)
gauge field, leaving the other polarisation massless.

The presence of the UV/IR effects in the trace-U(1) factors
makes it pretty clear that a simple
noncommutative U(1) theory taken on its own has nothing to do with ordinary QED.
The low-energy theory emerging from
the noncommutative U(1) theory will become free at $k^2 \to 0$ (rather than just
weakly coupled) and in addition will have other pathologies
~\cite{Khoze:2004zc,Khoze:2000sy,Hollowood:2001ng,Alvarez-Gaume:2003}.
However,
one would expect that it is conceivable to embed a
commutative $\textrm{SU}(N)$ theory, such as e.g. QCD or the weak sector of the Standard Model
into a supersymmetric noncommutative theory in the UV, but some extra care should be
taken with the QED U(1) sector~\cite{Khoze:2004zc}. We will show that
the only realistic way to embed QED into
noncommutative settings is to recover the electromagnetic U(1) from
a {\it traceless} diagonal generator of some higher $\textrm{U}(N)$ gauge theory.
So it seems that in order to embed QED into a noncommutative theory one should
learn how to embed the whole Standard Model~\cite{Khoze:2004zc}. We will see, however,
that the additional trace-U(1) factors remaining from the noncommutative
$\textrm{U}(N)$ groups will make the resulting low-energy theories unviable
(at least for the general class of models considered in this paper).

In order to proceed we would like to disentangle the mass-effects due to the Higgs mechanism
from the mass-effects due to non-vanishing $\Pi_2.$ Hence we first set $\Pi_2= 0$
(this can be achieved by starting with an exactly supersymmetric theory).
It is then straightforward to show (see Sec. \ref{prove}) that the Higgs mechanism alone cannot
remove all of the trace-U(1) factors from the massless theory.
More precisely, the following statement is true:
{\it Consider a scenario where a set of fundamental, bifundamental and adjoint Higgs fields breaks
$\textrm{U}(N_1)\times \textrm{U}(N_2)\times\cdots \times \textrm{U}(N_m) \rightarrow H,$
such that $H$  is non-trivial.
Then there is at least one generator of the unbroken subgroup $H$ with {\it non-vanishing trace}. This generator can be chosen such that it generates a U(1) subgroup.}

We can now count all the massless U(1) factors in a generic noncommutative theory
with $\Pi_2= 0$ and after the Higgs symmetry breaking. In general we can have the following scenarios
for massless U(1) degrees of freedom in $H$:
\begin{enumerate}[(a)\,]
\item{}\label{possa}$\textrm{U}(1)_{Y}$ is traceless and in addition there
is one or more factors of trace-U(1) in $H$.
\item{}\label{possb} $\textrm{U}(1)_{Y}$ arises from a mixture of traceless and trace-U(1) generators of the
noncommutative product group $\textrm{U}(N_1)\times \textrm{U}(N_2)\times\cdots \times \textrm{U}(N_m).$
\item{}\label{possc} $\textrm{U}(1)_{Y}$ has an admixture of trace-U(1) generators as in (\ref{possb})
plus there are additional massless
trace-U(1) factors in $H$.
\end{enumerate}

In the following sections we will see that none of these options lead to an acceptable
low-energy theory once we have switched on $\Pi_2 \neq 0$, i.e. once we have introduced
mass differences between superpartners.
It is well-known \cite{Matusis:2000jf,Alvarez-Gaume:2003}
that
$\Pi_2 \neq 0$ leads to strong Lorentz symmetry violating effects in the dispersion relation
of the corresponding trace-U(1) vector bosons, and in particular, to mass-difference
of their helicity components. If option (\ref{possa}) was realised in nature, it would lead
(in addition to the standard photon) to a new colourless vector field with one polarisation
being massless, and one massive due to $\Pi_2.$

The options (b) and (c) are also not viable since an admixture of the
trace-U(1) generators to the photon would also perversely affect
photon polarisations and make some of them massive\footnote{One could hope that the
trace-U(1) factors could be made massive at the string scale
by working in a theory where these factors are anomalous. Then one could use the
Green-Schwarz mechanism \cite{Green:1984sg} to cancel the anomaly and simultaneously give a large stringy mass to these U(1) factors.
This scenario which is often appealed to in ordinary commutative theories to remove
unwanted U(1) factors cannot be used in the noncommutative setting.
The reason is that at scales above the noncommutative mass, the noncommutative gauge invariance
requires the gauge group to be U($N$). It cannot become just an SU($N$) theory
(above the noncommutative scale) and remain noncommutative, see e.g. \cite{Armoni:2002fh}. Therefore we require vector-like theories as stated in item 5.}.

In the rest of the paper we will explain these observations in more detail.

We end this section with some general comments on noncommutative Standard Modelling.
This paper refines the earlier analysis
of \cite{Khoze:2004zc}. In that work the trace-U(1) factors were
assumed to be completely decoupled in the extreme infrared and, hence, were
neglected. However, it is important to keep in mind that the decoupling
of the trace-U(1)'s is
logarithmic and hence slow. Even
in presence of a huge hierarchy between the noncommutative mass scale $M_{\textrm{NC}}$,
say of the order of the Planck scale
$M_{\textrm{P}}\sim 10^{19}\ \textrm{GeV}$, and the scale
$\Lambda \sim (10^{-14}-10^{9})\,\textrm{eV}$ (electroweak and QCD scale, respectively),
where the SU($N$) subgroup becomes strong, the ratio
\begin{equation}
\label{ration1}
\frac{g^2_{\rm U(1)}}{g^2_{\textrm{SU}(N)}}\sim \frac{\log\left(\frac{k^2}
{\Lambda^{2}}\right)}{\log\left(\frac{M^4_{\textrm{NC}}}{\Lambda^{2}k^2}\right)}
\gtrsim 10^{-3}\, 
\end{equation}
is not negligible. In particular, the above inequality holds for any $M_{\textrm{NC}}>k\gtrsim 2\Lambda$. 
Hence the complete decoupling of the trace-U(1) degrees of freedom at small non-zero
momenta does not appear to be fully justified and the trace-U(1) would leave its traces in scattering experiments at accessible momentum scales $k\sim 1\,\textrm{eV}-10^{10}\,\textrm{eV}$ (see Sec. 2 for more detail).

\section{UV/IR mixing and properties of the trace-U(1)}\label{example}
UV/IR mixing manifests itself only in the trace-U(1) part of the full noncommutative U($N$).
For this part it strongly affects
$\Pi_{1}$ and is responsible for the generation of nonvanishing $\Pi_{2}$ (if SUSY is not exact).
In this section we will briefly
review how the UV/IR mixing arises in the trace-U(1) sector and how this leads us to rule out
options (a) and (c) discussed in Sec.~1.

\subsection{Running gauge coupling}
Following Refs.~\cite{Khoze:2000sy,Hollowood:2001ng},
we will consider a U($N$) noncommutative theory with matter
fields transforming in the adjoint and fundamental representations of the gauge group.
We use the background field method,
decomposing the gauge field $A_\mu = B_\mu + N_\mu$ into a background field $B_\mu$ and a
fluctuating quantum field $N_\mu$,
and the appropriate background version of Feynman gauge,
to determine the effective action $S_{\rm eff}(B)$ by functionally integrating over the fluctuating
fields.

To determine the effective gauge coupling in the background field method,
it suffices to study the terms quadratic in the background field. In the effective action these take
the following form (capital letters denote full U($N$) indices and run from $0$ to $N^{2}-1$)
\footnote{We use euclidean momenta when appropriate and the analytic continuation when considering
the equations of motion in subsection \ref{eom}.},
\begin{equation}
S_{\rm eff}   \ni
2\int \frac{d^{4}k}{(2\pi)^4} B^{A}_{\mu}(k)B^{B}_{\nu}(-k)\Pi^{AB}_{\mu\nu}(k).
\end{equation}
At tree level, $\Pi^{AB}_{\mu\nu}=(k^2 g_{\mu\nu}-k_{\mu}k_{\nu})\,\delta^{AB}/g^2_0$ is the standard
transverse tensor originating from the gauge kinetic term. In a commutative theory, gauge and Lorentz
invariance restrict the Lorentz structure to be identical to the one of the tree level term.
In noncommutative theories, Lorentz invariance is violated by
$\theta$. The most general allowed structure is then given by Eq. \eqref{poltensor}.
The second term may lead to the strong Lorentz violation mentioned in the introduction.
This term is absent in supersymmetric theories \cite{Matusis:2000jf,Khoze:2000sy}.

Let us start with a discussion of the effects noncommutativity has on $\Pi_{1}$ and the running of the gauge coupling.
That is, for the moment, we postpone the study of $\Pi_2$-effects by considering a
model with unbroken supersymmetry\footnote{Nevertheless, we will
give general expressions for $\Pi_{1}$ valid also in the non-supersymmetric case.}.
As usual, we define the running gauge coupling as
\begin{equation}
\label{defcoupling}
\left(\frac{1}{g^{2}}\right)^{AB}=\left(\frac{1}{g^{2}_{0}}\right)^{AB}+4\Pi^{AB}_{1\,\,\textrm{loop}}(k).
\end{equation}
where $g^{2}_{0}$ is the microscopic coupling (i.e. the tree level contribution) and
$\Pi_{\textrm{loop}}$ includes only the contributions from loop diagrams.
Henceforth, we will drop the loop subscript.

To evaluate $\Pi$ at one loop order one has to evaluate the appropriate Feynman diagrams.
The effects of noncommutativity appear
via additional phase factors $\sim \exp(i \frac{p \tilde{k}}{2})$ in the loop-integrals.
Using trigonometric relations one can group the integrals into terms where these factors combine
to unity, the so called planar parts, and those where they yield $\sim \cos ({p\tilde{k}})$,
the so called non-planar parts.

For fields in the fundamental representation, the phase factors cancel exactly\footnote{One may roughly
imagine that for each fundamental field
that appears in a Feynman diagram there is also the complex conjugate field which cancels the exponential factor.}
and only the
planar part is non-vanishing.
Fundamental fields therefore contribute as in the commutative theory \cite{Khoze:2000sy}.
In all loop integrals\footnote{To keep the equations simple we consider in this section a situation
where all particles of a given spin and
representation have equal diagonal masses.
Please note that the masses for fermions and bosons in the same representation may be different
as required for SUSY breaking.}
involving adjoint fields one finds the following factor \cite{Hollowood:2001ng},
\begin{equation}
M^{AB}(k, p) =
(-d \sin {k \tilde{p}\over 2}+ f \cos {k \tilde{p}\over 2})^{ALM}
(d \sin {k \tilde{p}\over 2}+ f \cos {k \tilde{p}\over 2})^{BML}.
\end{equation}
Using trigonometric and group theoretic relations this collapses to
\begin{equation}
M^{AB}(k,p) = - N \ \delta^{AB} (1-\delta_{0A}\cos k\tilde{p}).
\end{equation}
We can now easily see that all effects from UV/IR mixing, marked by the presence of the $\cos k\tilde{p}$,
appear only in the trace-U(1) part of the gauge group.
The planar parts, however, are equal for the U(1) and SU($N$) parts.

Summing everything up we find
the planar contribution (the coefficients $\alpha_{j},C_j,d_j$ are given in Table \ref{coefficients} and $C({\bf r})$
is the Casimir operator in the representation ${\bf r}$)
\begin{eqnarray}
\label{planarsusy2}
&&\Pi_{1\,\textrm{planar}} (k^2) =
-{2 \over (4\pi )^2 }\bigg( \sum_{j, {\bf r}} \alpha_{j}  C({\bf r})
\bigg[2C_j+\frac{8}{9}d_j
\\\nonumber
&&\quad\quad\quad\quad\quad\quad\quad\quad\quad\quad\quad\quad\quad+
\int_{0}^{1} dx \left(C_j-(1-2x)^2d_j\right)\ \log {A(k^2 , x,m^2_{j, {\bf r}}) \over \Lambda^2} \bigg]\bigg),
\end{eqnarray}
where $m_{j, {\bf r}}$ is the mass of a spin $j$ particle belonging to the representation $\bf r$ of the gauge group,
\begin{equation}
A(k^2,x,m^2_{j, {\bf r}})=k^2 x(1-x)+m^2_{j, {\bf r}},
\end{equation}
and
$\Lambda$  appears via dimensional transmutation similar to $\Lambda_{\overline{\textrm{MS}}}$ in QCD.
We have chosen the renormalisation scheme, i.e. the finite constants, such that $\Pi_{1\,\textrm{planar}}$ vanishes
at $k=\Lambda$.
\begin{table}[!t]
\begin{center}
\begin{tabular}{|c|c|c|c|c|}
\hline  j=& scalar  & Weyl fermion & gauge boson  & ghost \\
\hline $\alpha_{j}$ & -1 & $\frac{1}{2}$ & $-\frac{1}{2}$ &1  \\
\hline  $C_j$ & 0 &  $\frac{1}{2}$& 2 & 0 \\
\hline  $d_j$ & 1 & 2 & 4 & 1 \\
\hline
\end{tabular}
\end{center}
\caption{Coefficients appearing in the evaluation of the loop diagrams.}
\label{coefficients}
\end{table}

For the trace-U(1) part the nonplanar parts do not vanish and we find
\begin{equation}
\Pi_{1\,{\rm nonplanar}} = { 1\over 2 k^2}\left(\hat{\Pi} - \tilde{\Pi}\right) ,
\end{equation}
with
\begin{eqnarray}
\hat{\Pi} &=& {C({\bf G}) \over (4\pi)^2}\left\{
{8 d_j \over \tilde{k}^2} - k^2\left[ 12C_j - d_j\right]
\int_{0}^{1}dx \
K_{0} (\sqrt{A} |\tilde{k}|)\right\}
\ \ ,
\\
\tilde{\Pi}& =& {4C({\bf G})\over (4\pi)^2}\left\{
{ d_j \over \tilde{k}^2}-  \left(C_j k^2 -
d_j
{\partial^2 \over \partial^2 |\tilde{k}| }  \right)
 \int_{0}^{1}dx \
K_0 (\sqrt{A} |\tilde{k}|)\right\} ,
\end{eqnarray}
where $C({\bf G})=N$ is the Casimir operator in the adjoint representation.

\begin{figure}
\begin{center}
\scalebox{0.95}[0.95]{
\begin{picture}(190,180)(40,0)
\includegraphics[width=9.5cm]{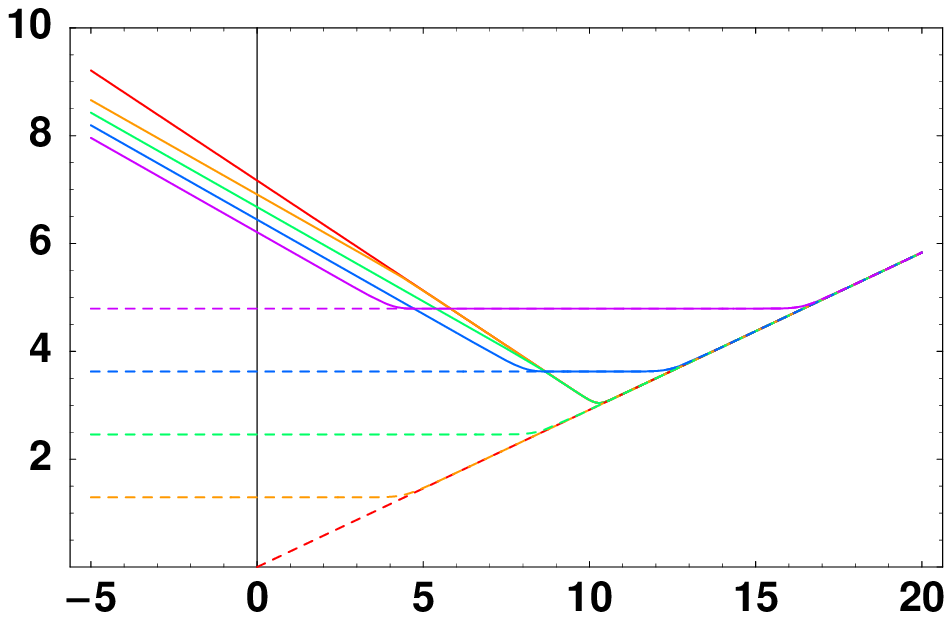}
\Text(-40,-15)[c]{\scalebox{1.2}[1.2]{$\log_{10}(k/\Lambda)$}}
\Text(-280,150)[c]{\scalebox{1.7}[1.7]{$\frac{1}{g^2}$}}
\end{picture}
}
\end{center}
\caption{The running gauge couplings $g_{\textrm{U(1)}}$ (solid) and $g_{\textrm{SU(2)}}$ (dashed)
for a U(2) theory with two matter
multiplets and all particles of equal mass
$m=0,10^4,10^8,10^{12},10^{16}\,\Lambda$, from top to bottom (left side, solid),
as a function of the momentum $k$, for
a choice of
$|\tilde{k}|=\theta_\textrm{eff} |k|$, with $\theta_{\textrm{eff}}=10^{-20}\Lambda^{-2}$.
}
\label{u1gaugecoupling}
\end{figure}
For illustration, we plot in Fig.~\ref{u1gaugecoupling}
the coupling~\eqref{defcoupling} for a toy model which is a supersymmetric U(2)
gauge theory with two matter multiplets and all masses (of all fields) taken to be equal.
We observe that even for large masses the running of the U(1) part (solid lines) does not stop in the infrared.
For masses smaller than the noncommutative mass scale $m^2\ll M_{\textrm{NC}}$  the trace-U(1)
gauge coupling has a sharp bend at $M_{\textrm{NC}}$ where the nonplanar parts
start to contribute. For larger masses the running stops at the mass scale $m^2$ only to resume running at a scale
$\sim M^{4}_{\textrm{NC}}/m^2$ which is, of course,
again due to the nonplanar parts.
The dashed lines in Fig.~\ref{u1gaugecoupling} give the running of the SU(2) part which receives
no nonplanar contributions and behaves like in an ordinary commutative theory.
For $m^2=0$ the SU(2) gauge coupling reaches a Landau pole at
$k=\Lambda$, for all non vanishing masses the running stops at the mass scale.
We observe that the ratio between the SU(2) coupling and the trace-U(1) coupling is not
incredibly small over a wide range of scales,
in support of our assertion~\eqref{ration1} in Sec. 1.

Further support comes from looking at the following approximate form for the running of the gauge coupling.
We assume the hierarchy
$\Lambda^{2}\ll m^2\ll M^{2}_{\textrm{NC}}$,
\begin{eqnarray}
\label{run1}
&&\!\!\!\!\!\frac{4\pi^2}{g^2_{\textrm{U(1)}}}=b^{\textrm{p}}_{0}\log\left(\frac{k^2}{\Lambda^2}\right),
\,\quad\quad\quad\quad\quad\quad\quad\quad\quad\quad\quad\quad\quad\quad\quad\quad\quad\,\,\textrm{for}\quad
k^2 \gg M^{2}_{\textrm{NC}},
\\\nonumber
&&\!\!\!\!\!\frac{4\pi^2}{g^2_{\textrm{U(1)}}}=b^{\textrm{p}}_{0}\log\left(\frac{k^2}{\Lambda^2}\right)
-b^{\textrm{np}}_{0} \log\left(\frac{k^2}{M^{2}_{\textrm{NC}}}\right),\,
\quad\quad\quad\quad\quad\quad\quad\quad\quad\,\,\, \textrm{for}\quad  m^{2}\ll k^2\ll
M^{2}_{\textrm{NC}},
\\\nonumber
&&\!\!\!\!\!\frac{4\pi^2}{g^2_{\textrm{U(1)}}}=b^{\textrm{p}}_{0}\log\left(\frac{m^2}{\Lambda^2}\right)
-b^{\textrm{np}}_{0} \left[\log\left(\frac{m^2}{M^{2}_{\textrm{NC}}}\right)
+\frac{1}{2}\log\left(\frac{k^2}{m^2}\right)\right],\,\quad\,\,\,\, \textrm{for}\quad  k^2 \ll m^2.
\end{eqnarray}
The gauge coupling for the SU($N$) subgroup $g^2_{\textrm{SU}(N)}$ is obtained by setting
$b^{\textrm{np}}_{0}=0$.
For simplicity let us now consider a situation where we have only fields
in the adjoint representation.
One finds \cite{Khoze:2004zc,Hollowood:2001ng} that $b^{\textrm{np}}_{0}=2b^{\textrm{p}}_{0}$, and
\begin{eqnarray}
\label{run2}
&&\frac{g^{2}_{\textrm{U}(1)}}{g^{2}_{\textrm{SU}(N)}}=1,
\quad\quad\quad\quad\quad\quad \textrm{for} \quad k^2\gg M^{2}_{\textrm{NC}},
\\\nonumber
&&\frac{g^{2}_{\textrm{U}(1)}}{g^{2}_{\textrm{SU}(N)}}
=\frac{\log\left(\frac{k^2}{\Lambda^{2}}\right)}{\log\left(\frac{M^{4}_{\textrm{NC}}}{\Lambda^{2}k^2}\right)},
\quad\,\,\,\,\, \textrm{for}\quad m^{2}\ll k^2\ll M^{2}_{\textrm{NC}},
\\\nonumber
&&\frac{g^{2}_{\textrm{U}(1)}}{g^{2}_{\textrm{SU}(N)}}
=\frac{\log\left(\frac{m^2}{\Lambda^{2}}\right)}{\log\left(\frac{M^{4}_{\textrm{NC}}}{\Lambda^{2}k^2}\right)},
\quad\,\,\,\,\, \textrm{for} \quad k^{2}\ll m^2.
\end{eqnarray}
To reach
\begin{equation}
\frac{g^{2}_{\textrm{U}(1)}}{g^{2}_{\textrm{SU}(N)}}<\epsilon = 10^{-3}
\end{equation}
we need $\log\left(\frac{M^{4}_{\textrm{NC}}}{\Lambda^{2}k^2}\right)$ and in turn $M_{\textrm{NC}}$ to be large.

As a generic example let
us use $\Lambda=\Lambda_{W}\sim 10^{-14}\,\textrm{eV}$ (the scale where the ordinary electroweak
SU(2) would become strong, in absence of electroweak symmetry breaking)
and $k=1\,\textrm{eV}$\footnote{It is obvious that $k^2\ll M^{2}_{\textrm{NC}}$. In this regime our
formulas \eqref{run1} and \eqref{run2} approximate the full result to a very high precision since threshold effects are negligible.}. We find
\begin{equation}
\label{abschaetz}
M_{\textrm{NC}}>
\Lambda^{\frac{1}{2}}k^{\frac{1}{2}}\exp\left(\frac{1}{4\epsilon}\log\left(\frac{k^2}{\Lambda^2}\right)\right)
\sim 10^{6974}\,M_{\textrm{P}}.
\end{equation}
Taking electroweak symmetry breaking into account we have to replace $\log\left(\frac{k^2}{\Lambda^2}\right)$
by $\log\left(\frac{M^2_{\textrm{EW}}}{\Lambda^2}\right)$ with $M_{\textrm{EW}}\sim 100\,\textrm{GeV}$
in \eqref{abschaetz}. We find
\begin{equation}
M_{\textrm{NC}}>10^{12474}\,M_{\textrm{P}}.
\end{equation}
Let us increase the coupling strength of the SU($N$) by using $\Lambda=0.5\,\textrm{eV}$. $k=1\,\textrm{eV}$
is now quite close to the strong coupling scale of the SU($N$). Without symmetry breaking we find
\begin{equation}
M_{\textrm{NC}}>10^{131}\,M_{\textrm{P}}.
\end{equation}
We might be able to reduce this number by some orders of magnitude but without using an extreme field content
it remains always incredibly large. Indeed, one can typically find a scale $k$ which is not too close to the strong coupling scale of the SU($N$) which strengthens the bounds dramatically. Therefore, as a conservative estimate we
propose\footnote{Of course, this constraint should not be taken overly serious. Above the string scale one should perform
a string theory analysis. The main point is that the scale we find is way beyond the Planck scale.}
\begin{equation}
M_{\textrm{NC}}>10^{100}\,M_{\textrm{P}}.
\end{equation}

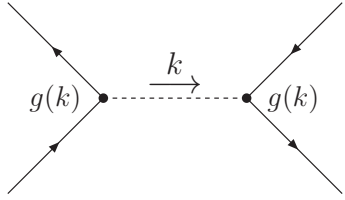
\begin{figure}[t]
\begin{center}
\scalebox{0.9}[0.9]{
\begin{picture}(190,140)(0,0)
\SetOffset(3,10)
\ArrowLine(40,60)(0,100)
\ArrowLine(0,20)(40,60)
\DashLine(40,60)(100,60){2}
\ArrowLine(140,100)(100,60)
\ArrowLine(100,60)(140,20)
\Vertex(40,60){2}
\Text(20,60)[c]{\scalebox{1.0}[1.0]{$g(k)$}}
\Text(120,60)[c]{\scalebox{1.0}[1.0]{$g(k)$}}
\Vertex(100,60){2}
\Text(70,75)[c]{\scalebox{1.1}[1.1]{$k$}}
\Text(70,65)[c]{\scalebox{1.2}[1.2]{$\longrightarrow$}}
\end{picture}}
\end{center}
\vspace{-1.0cm}
\caption{A typical Feynman diagram for scattering. The effective coupling $g$ depends on the momentum $k$.}
\label{scattering}
\end{figure}

To conclude this subsection, let us point out that, in a scattering experiment (as depicted in Fig. \ref{scattering}),
$k$ is really the scale of the internal momentum, and therefore, non-vanishing.
$\tilde{k}$, too, is non-vanishing in appropriate (remember that we have Lorentz symmery violation)
directions of $t$-channel scattering.

\subsection{The effects of a non vanishing $\Pi_{2}$ from SUSY breaking}\label{eom}

In the previous subsection we made
$\Pi_{2}$ vanish by working in a supersymmetric theory.
Let us now study, what happens, when supersymmetry is (softly) broken.

Looking only at the trace-U(1) degrees of freedom of a generic noncommutative theory we have
\begin{equation}
\Pi_{2}=\sum_{j}\alpha_{j}\left[\frac{1}{2}(3\tilde{\Pi}_{j}-\hat{\Pi}_{j})\right].
\end{equation}
One easily checks that
\begin{equation}
\Pi_{2}\sim \sum_{j} \alpha_{j} d_j f(k^2,\tilde{k}^2,m_j).
\end{equation}
If SUSY is unbroken, all masses are equal. Using supersymmetric matching between
bosonic and fermionic degrees of freedom,
\begin{equation} \label{susycanc}
\sum_{j} \alpha_{j} d_j=0,
\end{equation}
we reproduce the vanishing of $\Pi_{2}$.
If SUSY is softly broken this cancellation is not complete anymore
(in fact \eqref{susycanc} still holds and this removes the leading power-like IR divergence
in $\Pi_{2}$, however, the subleading effects in $\Pi_{2}$ survive).
$\Pi_{2}$
gets a contribution \cite{Alvarez-Gaume:2003}
\begin{eqnarray}
\label{pi2}
\Pi_{2}\!\! &=& \!\!D\sum_{j}\alpha_{j}d_jm^{2}_{j}\left[K_{0}(m\tilde{k})+K_{2}(m\tilde{k})\right]+O(k^2)
\\\nonumber
\!\! &=& \!\! C \Delta M^2_{\textrm{SUSY}}+C'\sum_{j}\alpha_{j}d_j m^{2}_{j}\log(m^2_{j}\tilde{k}^2)+\cdots,
\end{eqnarray}
with known constants $C$, $C'$ and $D$.
This has dire consequences for the gauge boson. Let us look at the equations of motion resulting from this
additional Lorentz symmetry violating contribution to the polarisation tensor (we briefly review the equations of motion
for ordinary photons in Appendix \ref{polarisation}).

In presence of a Higgs field which generates a mass term $m^2$ and using unitary gauge the field
equations in presence of non vanishing $\Pi_2$ read
\begin{equation}
\label{fieldnc} \left(\Pi_{1}(k^{2}g_{\mu\nu}-k_{\mu}k_{\nu})
+\Pi_{2}\frac{\tilde{k}_{\mu}\tilde{k}_{\nu}}{\tilde{k}^{2}}-m^{2}g_{\mu\nu}\right)A^{\nu}=0.
\end{equation}
Using that unitary gauge implies Lorentz gauge,
$k_{\mu}A^{\mu}=0$, we can simplify
\begin{equation}
\label{fieldnc2}
(\Pi_{1}k^{2}-m^{2})A_{\mu}+\Pi_{2}\frac{\tilde{k}_{\mu}\tilde{k}_{\nu}}{\tilde{k}^{2}}A^{\nu}=0.
\end{equation}
To proceed further it is useful to specify a direction for the
momentum and the noncommutativity parameters. The photon flies in
3-direction and we have
\begin{equation}
k^{\mu}=(k^{0},0,0,k^{3}).
\end{equation}
What is the corresponding value of $\tilde k$?
Since $\theta^{\mu\nu}$ breaks Lorentz invariance, we
need to specify $\theta^{\mu\nu}$ in a particular frame.
For the latter, a natural one is the system where the cosmic microwave background is at rest.
In this frame, we assume that the only non-vanishing components of $\theta^{\mu\nu}$ are
\begin{equation}
\theta^{13}=-\theta^{31}=\theta .
\end{equation}

This yields,
\begin{equation}
\tilde{k}_{\mu}=\theta_{\mu\nu}k^{\nu}=(0,\theta k^{3},0,0),\quad
\tilde{k}^{\, 2}=(\theta k^{3})^{2}.
\end{equation}
We start with the ordinary transverse components of $A^{\nu}$,
\begin{equation}
A^{\nu}_{1}=(0,1,0,0).
\end{equation}
In this direction, \eqref{fieldnc2} yields
\begin{equation}
\label{direction1}
(\Pi_{1}k^{2}-m^{2}-\Pi_{2})A_{1,\nu}=0.
\end{equation}
In the other transverse direction,
\begin{equation}
A^{\mu}_{2}=(0,0,1,0),
\end{equation}
we find
\begin{equation}
(\Pi_{1}k^{2}-m^{2})A_{2,\nu}.
\end{equation}
Finally we have the third polarisation (which can be gauged away if and only if $m^2=0$),
\begin{equation}
A^{\mu}=(a,0,0,b),\quad k^{0}a-k^{3}b=0
\end{equation}
which results in
\begin{equation}
(\Pi_{1}k^{2}-m^{2})A_{3,\nu}.
\end{equation}
We note that the different polarisation states do not
mix due to the presence of $\Pi_{2}$. The second and the third polarisation state behave more
or less like in the ordinary commutative case. However, the first has a modified
equation of motion, \eqref{direction1},
in presence of a non-vanishing $\Pi_{2}$\footnote{One might argue that instead of
Eq. \eqref{direction1} one has to use the rescaled equation (we set $m^2=0$ for
simplicity)
$k^2-\frac{\Pi_{2}(k^2,\tilde{k}^2)}{\Pi_{1}(k^2,\tilde{k^2})}=0$.
For $k^2\to 0$, the second term vanishes since $\Pi_{1}$ diverges in this limit.
Therefore, we find an additional solution.
However, this solution is rather strange. It does not correspond to a pole in the
propagator (it goes like a $\log$).
Moreover, if one calculates the cross section $\Pi_{2}$ still upsets the angular
dependence quite severely compared to
the ordinary commutative case.}.

This is another strong argument against a trace-U(1) being the photon \cite{Alvarez-Gaume:2003}.
If the gauge symmetry is unbroken and $m^2=0$ we usually have two massless polarisations.
However, a non vanishing $\Pi_{2}$ reduces this to one. The other one gets an additional mass $\Pi_{2}$.
Since only one polarisation is affected this is a strong Lorentz symmetry violating effect.
Moreover, a negative $\Pi_{2}$ would lead to tachyons while a positive mass is phenomenologically ruled
out by the constraint \cite{Eidelman:2004wy}
\begin{equation}
m_{\gamma}<6\times 10^{-17}\,\textrm{eV}
\end{equation}
on the photon mass\footnote{Even fine-tuning of \eqref{pi2} to zero is not an option.
Since we have only a finite number of masses this is at best possible for a finite number of values
of $|\tilde{k}|$ and we will surely find values of $|\tilde{k}|$ where $\Pi_{2}$ is nonzero.}.

If we take the trace-U(1) as an additional (to the photon)
gauge boson from the unbroken subgroup $H$, we would still get strong Lorentz symmetry
violation since the trace-U(1) is not completely decoupled.

In summary, we found in this section that additional trace-U(1) subgroups
are not completely decoupled and should lead to observable effects.
In particular, if SUSY is not exact we have non-vanishing $\Pi_{2}$
which gives rise to strong Lorentz symmetry violation which has not been observed.
This rules out possibilities (\ref{possa}) and (\ref{possc}) of Sec.~1.
Moreover, we confirmed that a trace-U(1) is not suitable as a photon candidate.

\section{Mixing of trace and traceless parts}\label{u2example}
{}From the previous section we concluded that the trace-U(1) groups are unviable
as candidates for the SM photon.
Therefore, it has been suggested to construct the photon from traceless U(1) subgroups \cite{Khoze:2004zc}.
It turns out, however, that typically trace and traceless parts mix and the trace parts
contribute their Lorentz symmetry violating properties
to the mixed particle.

For U(2) broken by a fundamental Higgs, the standard Higgs mechanism yields the symmetry breaking $U(2)\to U(1)$.
However, the remaining U(1) is a mixture of trace and traceless parts.
If SUSY is broken, the trace-U(1) has a $\Pi_{2}$ part in the polarisation tensor.
Taking this into account we find the following matrix for the equations of motion
\begin{eqnarray}
\label{final}
\begin{tiny}
\left(\begin{array}{cccccc}
 \Pi^{\textrm{U(1)}}_{1}k^{2}-\Pi_{2}-m^{2} &  m^2 &  &  &  &  \\
m^2 & \Pi^{\textrm{SU(2)}}_{1}k^2-m^{2} &  &  &  &  \\
 &  & \Pi^{\textrm{U(1)}}_{1}k^{2}-m^{2} & m^2 &  &  \\
 &  &  m^2 &\Pi^{\textrm{SU(2)}}_{1}k^2-m^{2}   &  &  \\
 &  &  &  & \Pi^{\textrm{U(1)}}_{1}k^2-m^{2} &m^{2}  \\
 &  &  &  & m^{2} &\Pi^{\textrm{SU(2)}}_{1}k^2-m^{2}
\end{array}  \right),
\end{tiny}
\end{eqnarray}
where the adjoint U(2) and polarisation indices are $(0,1),(3,1),(0,2),(3,2),(0,3),(3,3)$.
We omitted the values $1$ and $2$ for the adjoint U(2) indices which do not mix with
the trace-U(1) and are not qualitatively different from the commutative case.

The matrix is block
diagonal and the second and third polarisation (lower right corner)
behave more or less like their commutative counterparts.
We can concentrate on the upper left $2\times2$ matrix corresponding
to the transverse polarisations affected
by $\Pi_{2}$.

This $2\times2$ matrix admits two solutions for the equations of motion. Expanding for small $\Pi_{2}$ we
find,
\begin{eqnarray}
\label{u2solution}
\left(\Pi^{\textrm{U(1)}}_{1}+\Pi^{\textrm{SU}(N)}_{1}\right)k^2\!\!&=&\!\!\Pi_{2}+O(\Pi^{2}_{2}),
\\\nonumber
\left(\Pi^{\textrm{U(1)}}_{1}+\Pi^{\textrm{SU}(N)}_{1}\right)k^2\!\!&=&\!\!\
\frac{\left(\Pi^{\textrm{U(1)}}_{1}+\Pi^{\textrm{SU}(N)}_{1}\right)^2}{\Pi^{\textrm{U}(1)}_{1}\Pi^{\textrm{SU}(N)}_{1}} m^2 +\frac{\Pi^{\textrm{SU}(N)}_{1}}{\Pi^{\textrm{U}(1)}_{1}}\Pi_{2}+O(\Pi^{2}_{2}),
\end{eqnarray}
in analogy to \eqref{direction1}. In absence of $\Pi_{2}$ the first solution in
Eq. \eqref{u2solution} is a massless one
corresponding to the massless combination of gauge bosons (think of it as the photon).
The second is a massive combination (similar to the $Z$ boson). The presence of non-vanishing $\Pi_{2}$
again leads to a mass $\frac{\Pi_{2}}{\Pi^{\textrm{U(1)}}}$ for the first
solution and rules out the ``massless'' combination
as a reasonable photon candidate.

This example demonstrates that the disastrous effects of $\Pi_{2}$ are also present in any combination
which has an admixture of trace-U(1) degrees of freedom.
Hence, this rules out possibilities (\ref{possb}) and (\ref{possc}) from the introduction.

\section{Trace-U(1) factors in the unbroken subgroup}\label{prove}
In the previous section, we learned in a specific example that even a small admixture of a
trace part spoils the masslessness of the gauge boson corresponding to the unbroken gauge symmetry.
This shows that a viable photon candidate must have a generator with vanishing (small is not enough)
trace.

In our U(2) example with the gauge symmetry broken by a fundamental Higgs field the trace does not vanish.
The generator corresponding to the unbroken U(1) is
\begin{equation}
\sim \left(
\begin{array}{cc}
1 & 0 \\
0 & 0
\end{array}
\right),
\end{equation}
which obviously has non-vanishing trace.

One can try to construct other symmetry breaking mechanisms with larger groups and products of groups as well
as the other representations for the Higgs fields allowed by the condition 3 of the introduction. However, one
always encounters one of the following situations. Either the remaining U(1) has a generator
with non-vanishing trace
or there is more than one unbroken U(1) subgroup.
Both situations are in contradiction of observations, as our discussion of the previous sections shows.

This is generalised and more precisely formulated by the following proposition (already stated in the introduction):
{\it Consider a scenario where a set of fundamental, bifundamental and adjoint Higgs fields breaks
$\textrm{U}(N_1)\times \textrm{U}(N_2)\times\cdots \times \textrm{U}(N_m) \rightarrow H,$
such that $H$  is non-trivial.
Then there is at least one generator of the unbroken subgroup $H$ with {\it non-vanishing trace}.
This generator can be chosen such that it generates a U(1) subgroup.}

Let us now turn to a proof of the proposition. Let us start with the simple situation of one U($N$) group.
Since we have only one group, we have only fundamental and adjoint Higgs
fields at our disposal.
We proceed by switching on one Higgs field (component) after the other.
Let us start with the fundamental field. U($N$) symmetry allows us to chose this field  as
\begin{equation}
\label{fundamental}
\phi_{\textrm{f}}=(0,\ldots,0,a)^{\textrm{T}}.
\end{equation}
{\it{Case 1:}} If $a=0$ we have no breaking with a fundamental Higgs. In this case we are finished, because the
generator of the original trace-U(1) is proportional to the $N\times N$ unit matrix and therefore commutes
with any adjoint Higgs field. Therefore this generator continues to generate an unbroken trace-U(1)
subgroup, as stated in the proposition.

\noindent{\it{Case 2:}} If $a\neq0$ gauge symmetry is broken down to the U($N-1$) living in the upper $N-1$
components of any field. A set of generators for this group are the ordinary U($N-1$) in the upper
left $(N-1)\times (N-1)$ submatrix and zero in the other components. In particular, there is a new trace-U(1)
with generator
\begin{equation}
T^{1}_{\textrm{trace}}=
\left(\begin{array}{cccc}
1 &  &  &  \\
 & \ddots &  &  \\
 &  & 1 &  \\
 &  &  & 0
\end{array}
\right).
\end{equation}
 Under this subgroup an adjoint field decomposes into
\begin{equation}
\label{adjoint}
\phi_{\textrm{ad}}=
\left(\begin{array}{c|c}
\phi^{2}_{\textrm{ad}}  & \phi^{2}_{\textrm{f}}\\
\hline  (\phi^{2}_{\textrm{f}})^{\dagger} &\phi^{2}_{\textrm{s}}  \\
\end{array}\right),
\end{equation}
where $\phi^{2}_{\textrm{ad}}$,  $\phi^{2}_{\textrm{f}}$ and $\phi^{2}_{\textrm{s}}$ are adjoint,
fundamental and singlett fields under
the remaining U($N-1$) symmetry.
An additional fundamental field $\hat{\phi}_{\textrm{f}}$ decomposes as
\begin{equation}
\label{fundamental}
\hat{\phi}_{\textrm{f}}=\left(
\begin{array}{c}
\hat{\phi}^{2}_{\textrm{f}} \\
\hat{\phi}^{2}_{\textrm{s}}
\end{array} \right)
\end{equation}
into an additional fundamental $\hat{\phi}^{2}_{\textrm{f}}$ and another singlett $\hat{\phi}^{2}_{\textrm{s}}$.
We can now repeat the argument for the remaining U($N-1$) group starting, again, with the fundamental fields.

\noindent This procedure has to stop at some point, i.e. at one point the fundamental $\phi^{n}_{\textrm{f}}$
has to be zero, or the
symmetry is broken completely and $H$ would be the trivial group in violation of the assumptions.

For a product of more than one group the proof is analogous only that we have
additional bifundamental fields. Let us briefly consider the situation with a product of two groups
$\textrm{U}(M)\times \textrm{U}(N)$. Switching on fundamental fields we can end up with:

\noindent {\it{Case 1:}} If all fundamentals are zero the symmetry remains unbroken
$\textrm{U}(M)\times \textrm{U}(N)$. One can easily see that bifundamental and adjoint fields cannot break the
trace-U(1) generated by the $(N+M)\times (N+M)$ unit matrix\footnote{We can think
of $\textrm{U}(M)\times \textrm{U}(N)$ embedded
into $\textrm{U}(N+M)\times\textrm{U}(N+M)$}.

\noindent {\it{Case 2:}} Let us switch on one fundamental field. Without loss of generality we can take it to be an
$N$ fundamental. The symmetry is broken down to $\textrm{U}(M)\times \textrm{U}(N-1)$ with a
new trace-U(1) for the U($N-1$) in
analogy to the simple U($N$) situation discussed above.
All fields transforming under the U($M$) remain unaffected. The fundamental and
adjoint fields for U($N$) are decomposed
according to Eqs. \eqref{adjoint},\eqref{fundamental}. Finally the bifundamental decomposes as
\begin{equation}
\phi_{b}=
\left(\begin{array}{c|c}
 \phi^{2}_{\textrm{b}} & \phi^{2}_{\textrm{b, f}}  \\
\end{array} \right)
\end{equation}
into a bifundamental $\phi^{2}_{\textrm{b}}$ under $\textrm{U}(M)\times \textrm{U}(N-1)$ and a
fundamental $\phi^{2}_{\textrm{b, f}}$ under U($N-1$).

\noindent The argument proceeds by induction. The case of more than two U($N$) factors is completely analogous.

\section{Conclusions}\label{conclusions}
Noncommutative gauge symmetry
in the Weyl-Moyal approach
leads to two main features which have to be taken into account for sensible
model building. First, there are strong constraints on the dynamics and the field content.
The only allowed gauge groups are U($N$).
In addition, the matter fields are restricted to transform as fundamental, bifundamental and adjoint
representations of the gauge group. Finally, anomaly freedom for noncommutative theories requires
the theory to be vector like\footnote{In turn, this eliminates the Green-Schwarz mechanism
\cite{Green:1984sg} as a possible
source for a (large) mass term for the trace-U(1) part of the gauge group.}.
Second, there are the effects of ultraviolet/infrared mixing.
Those lead to asymptotic infrared freedom of
the trace-U(1) subgroup and, if the model does not have unbroken supersymmetry, to Lorentz symmetry violating terms
in the polarisation tensor for this trace-U(1) subgroup.

We have demonstrated that, although the trace-U(1) decouples in the limit $k\to 0$, the coupling is not negligibly small
at finite momentum scales $k$, as they appear, for example, in scattering experiments. Therefore,
observations rule out additional unbroken (massless) trace-U(1) subgroups.
An example is the model considered in Ref. \cite{Khoze:2004zc}. In Ref. \cite{Khoze:2004zc}, the trace-U(1) groups
were completely discarded before the symmetry breaking scheme was discussed. A more careful
investigation which takes takes into account these subgroups yields the symmetry breaking
$\textrm{U}(4)\times \textrm{U}(3)\times \textrm{U} (2) \to \textrm{SU}(3)\times\textrm{SU}(2)\times (\textrm{U}(1))^4$ instead of
$\textrm{U}(4)\times \textrm{U}(3)\times \textrm{U} (2) \to \textrm{SU}(3)\times\textrm{SU}(2)\times \textrm{U}(1)$. Therefore we have
superfluous U(1) subgroups. Following the above lines explicitely one easily finds that one of the $\textrm{U}(1)$'s
has a generator which is proportional to the $9\times 9$ unity matrix.

Noncommutativity explicitly breaks Lorentz invariance. Therefore an additional \linebreak
Lorentz symmetry violating
structure is allowed in the polarisation tensor. This structure is absent only in
supersymmetric models. If supersymmetry is (softly) broken, this additional structure is present
in the polarisation tensor of the trace-U(1).
It leads to an additional mass $\sim \Delta M^2_{\textrm{SUSY}}$ for one of the transverse polarisation states
\cite{Alvarez-Gaume:2003}.
The tight constraints on the photon mass therefore exclude trace-U(1)'s as a candidate for the photon.
It turns out that even a small admixture of a trace part to a traceless part (unaffected by these problems) is fatal.
The only way out seems to be the construction of the photon from a completely traceless generator.
A group theoretic argument shows, that this is impossible whithout having additional
unbroken U(1) subgroups. However, those are already excluded from the arguments given above.

This result severely restricts the possibilities to construct a noncommutative Standard model extension.
If all of the constraints given at the beginning are fulfilled the noncommutativity scale is pushed
to scales far beyond $M_{\textrm{P}}$.
This is to be compared to the less restrictive constraints $M_{\textrm{NC}}\gtrsim 0.1-10\,\textrm{TeV}$
(conservative estimate) obtained from tree level amplitudes \cite{Hewett:2000zp}
or from an approach where a Taylor expansion in the noncommutativity parameters is used before
quantization, thereby ignoring effects of ultraviolet/infrared mixing and possibly constraints on the field content
\cite{Calmet:2001na,Madore:2000en,Carroll:2001ws,
Carlson:2001sw,Behr:2002wx,Schupp:2002up,Calmet:2004dn,Ohl:2004tn,Melic:2005su}.
We stress, however, that the latter approach may lead to a completely different quantum theory and therefore our bounds may not be applicable.

We would like to conclude with a more optimistic prospect.

In general there is no reason to assume that the simple noncommutative model
used here describes correctly the physics at energies ranging from a few eV up to the Planck mass.
In fact, due to the ultraviolet/infrared mixing, a different ultraviolet embedding of the theory would modify the theory
not only in the ultraviolet, but also in the infrared which can drastically alter our conclusions \cite{AJKR}.
In particular,
our conclusions are
tied to a slow logarithmic decoupling of the trace-U(1), but if it is changed to a power-like decoupling,
the U(1) factors would safely decouple and leave the Standard Model in peace.
We expect that this can be achieved by embedding the
noncommutative theory into a higher dimensional theory in the ultraviolet (which will have a power-like beta function)
and then appeal to the ultraviolet/infrared mixing to transport this power-like behaviour to the infrared region
for the trace-U(1) gauge coupling (see later work \cite{AJKR}).

\bigskip
\bigskip

\centerline{\bf Acknowledgements}

We would like to thank  Steve Abel and J\"urgen Reuter for useful discussions.
VVK acknowledges the support of PPARC through a Senior Fellowship.

\begin{appendix}
\section{Polarsation directions in gauge theories}\label{polarisation}
In this section we review some basics about the counting of degrees of freedom in gauge theories.
In particular, we show
how gauge invariance reduces the number of degrees of freedom from the naive 4 (4 components of the vector
field) to 2 and 3 for the massless and massive case, respectively.
\subsection{The massless case}
In ordinary QED, the field equations read
\begin{equation}
\label{field} \Box
A^{\mu}-\partial^{\mu}(\partial_{\nu}A^{\nu})=0.
\end{equation}
Using Lorentz gauge,
\begin{equation}
\label{lorentz}
\partial_{\mu}A^{\mu}=0
\end{equation}
Eq. \eqref{field} simplifies to the wave equation
\begin{equation}
\label{wave} \Box A^{\mu}=0.
\end{equation}
Writing
\begin{equation}
A^{\mu}=C \epsilon^{\mu} \exp(ikx),
\end{equation}
any $\epsilon^{\mu}$ is a solution to \eqref{wave} as long as
\begin{equation}
k^{2}=0.
\end{equation}
So far we have all $4$ polarisations. However, \eqref{lorentz}
implies 4-dimensional transversality,
\begin{equation}
\label{4dtransverse} k_{\mu}\epsilon^{\mu}=0,
\end{equation}
and reduces the allowed number of polarisations to three. This is
still more than the two polarisation states a photon should have.

However, Lorentz gauge does not completely fix the gauge. We can
still use a gauge transformation $\Omega$ with $\Box\Omega=0$.
This allows us to choose $A^{0}=0$. Together with
\eqref{4dtransverse} this leads us to the ordinary 3-dimensional
transversality,
\begin{equation}
\label{3dtransverse} \overrightarrow{k}\cdot \overrightarrow{\epsilon}=0.
\end{equation}
\subsection{The case with a Higgs field}
The presence of a Higgs field modifies \eqref{field},
\begin{equation}
\label{field2}
\Box A^{\mu}-\partial^{\mu}(\partial_{\nu}A^{\nu})+m^{2}A^{\mu}+m\partial^{\mu}\phi_{2}=0.
\end{equation}
Moreover it supplies an additional equation for the Goldstone
boson $\phi_{2}$,
\begin{equation}
\label{goldstone} \Box \phi_{2}+m\partial_{\mu}A^{\mu}=0.
\end{equation}
One convenient choice of gauge is unitary gauge where
\begin{equation}
\label{unitary} \phi_{2}=0.
\end{equation}
We stress from the beginning that unitary gauge implies
\eqref{lorentz}, as can be seen from \eqref{goldstone}. In this
gauge Eq. \eqref{field2} simplifies,
\begin{equation}
\label{wave2} \Box A^{\mu} +m^{2} A^{\mu}=0.
\end{equation}
Now everything runs in a similar fashion to the massless case,
only that
\begin{equation}
k^{2}-m^{2}=0.
\end{equation}
The important difference is that unitary gauge fixes
the gauge completely. We cannot make an additional gauge choice.
Therefore it is impossible to get rid of the 3rd polarisation
state which satisfies Lorentz gauge $k_{\mu}\epsilon^{\mu}=0$. Stated
differently we cannot require 3-dimensional transversality for
$\epsilon^{\mu}$ and we have therefore three allowed polarisation states
with equal masses.

\end{appendix}

\end{document}